\documentclass[prl,aps,superscriptaddress,twocolumn,notitlepage,showpacs]{revtex4-2}
\usepackage{latexsym}
\usepackage{amsmath}
\usepackage{amssymb}
\usepackage{graphicx}
\usepackage{caption}
\usepackage{subfigure}
\usepackage{float}
\usepackage{mathrsfs}
\usepackage{color}
\usepackage{txfonts}
\usepackage[justification=centering,format=plain]{caption}

\renewcommand{\raggedright}{\leftskip=0pt \rightskip=0pt plus 0cm}
\begin{document}

\title{Entangled Photons Enabled Time- and Frequency-Resolved Coherent Raman Spectroscopy in Condensed Phase Molecules}

%\title{Time- and Frequency-Resolved Coherent Raman Spectroscopy using Entangled Photons}

\author{Zhedong Zhang}
\email{zzhan26@cityu.edu.hk}
\affiliation{City University of Hong Kong, Kowloon Tong, Hong Kong SAR}

\author{Tao Peng}
\affiliation{Texas A$\&$M University, College Station, Texas 77843, United States}
%\affiliation{Institute for Quantum Science and Engineering, and Department of Physics and Astronomy, Texas A$\&$M University, College Station, Texas 77843, USA}

\author{Xiaoyu Nie}
\affiliation{Xi'an Jiaotong University, Xi'an, Shaanxi 710049, China}

\author{Girish S. Agarwal}
\affiliation{Texas A$\&$M University, College Station, Texas 77843, United States}
%\affiliation{Department of Biological and Agricultural Engineering, Texas A$\&$M University, College Station, Texas 77843, USA}

%\author{Shaul Mukamel}
%\affiliation{University of California Irvine, Irvine, California 92697, United States}

\author{Marlan O. Scully}
\affiliation{Texas A$\&$M University, College Station, Texas 77843, United States}
\affiliation{Baylor University, Waco, Texas 76704, United States}
\affiliation{Princeton University, New Jersey 08544, United States}
%\affiliation{Institute for Quantum Science and Engineering, and Department of Physics and Astronomy, Texas A$\&$M University, College Station, Texas 77843, USA}
%\affiliation{Quantum Optics Laboratory, Baylor Research and Innovation Collaborative, Waco, Texas 76704, USA}
%\affiliation{Department of Mechanical and Aerospace Engineering, Princeton University, New Jersey 08544, USA}

\date{\today}

\begin{abstract}
We develop an ultrafast frequency-resolved Raman spectroscopy with entangled photons for polyatomic molecules in condensed phases, to probe the electronic and vibrational coherences. Using quantum correlation between the photons, the signal shows the capability of both temporal and spectral resolutions that are not accessible by either classical pulses or the fields without entanglement. We develop a microscopic theory for this Raman spectroscopy, revealing the electronic coherence dynamics which often shows a rapid decay within $\sim 50$fs. The heterodyne-detected Raman signal is further developed to capture the phases of electronic coherence and emission in real-time domain.
%We develop a ultrafast frequency-resolved Raman spectroscopy with entangled photons for polyatomic molecules in condensed phases, to probe the electronic coherence. Using the quantum correlation between the photons, the signal shows the enhanced temporal and spectral resolutions that used to be conjugated in conventional Raman spectra. We develop a microscopic theory for this Raman spectroscopy, revealing the electronic coherence dynamics which often shows a rapid decay within $\sim 50$fs. The heterodyne-detected Raman signal is further developed to capture the phases of electronic coherence and emission in real-time domain. The paradigm of the signal is shown to be potentially a quantum extension of FAST CARS.
\end{abstract}

\maketitle

{\it Introduction}.--Over last few decades quantum entanglement has emerged to have considerable applications in testing fundamental laws of physics \cite{Mandel_PRL1987,Kwiat_PRL1995,Shih_PRL1988,Franson_PRL1989, Boyd_PRL2004,Zeilinger_PRL1998,Scully_PRL2000}, as well as revolutionizing photonic applications, e.g., quantum imaging \cite{Pittman_PRA1995}, metrology \cite{Shih_PRL2001,Mitchell_Nature2004}, sensing \cite{Cappellaro_RMP2017,Lloyd_NatPhotonics2018}, and quantum computing \cite{Zeilinger_Nature2005,OBrien_Science2007}.

%Quantum entanglement has draw much attention recently, as one of the most mysterious properties in quantum mechanics. Entanglement has been measured for decades by generating photon pairs through spontaneous parametric down-conversion (SPDC)~\cite{Kwiat_PRL1995}. %and has been used in the pioneering experiments to test Bell's inequality \cite{Shih_PRL1988,Franson_PRL1989, Boyd_PRL2004}. 
%The entangled photons hold great promise for testing fundamental laws of physics \cite{Shih_PRL1988,Franson_PRL1989, Boyd_PRL2004,Mandel_PRL1987,Zeilinger_PRL1998,Scully_PRL2000}, as well as revolutionizing photonic applications such as quantum imaging \cite{Pittman_PRA1995}, quantum metrology \cite{Shih_PRL2001,Mitchell_Nature2004}, quantum sensing \cite{Cappellaro_RMP2017,Lloyd_NatPhotonics2018}, and quantum computing \cite{Zeilinger_Nature2005,OBrien_Science2007}.

The field of spectroscopy especially the nonlinear spectroscopy can benefit enormously from the quantum states of light. The role of quantum entanglement in the context of two-photon spectroscopy was realized rather early. Fei {\it et al.} demonstrated entanglement-induced transparency in two-photon processes \cite{Fei_PRL1997}. The occurrence of simultaneous excitation two independent atoms due to entangled photons have been demonstrated in recent experiments \cite{Scully_PRL2004,Wang_PRL2020}. In general the entangled photons have demonstrated incredible power in developing new spectroscopic techniques for complex materials \cite{Mukamel_RMP2016,Mukamel_JPB2020,Raymer_JPCB2013,Oka_JCP2020}. Thanks to the technical advance in pulse shaping, the two-photon absorption with photon entanglement has been demonstrated in various experiments, incredibly suppressing the intermediate state relaxation that yields the optimal efficiency of populating higher excited states \cite{Mukamel_NatCommun2013,Goodson_JPCB2006,URen_PRL2019,Goodson_JPCL2017,Goodson_JACS2018}. The entangled states of photons provide extra control knobs for selective molecular relaxation and radiative processes. The related signals may be extended into sub-picosecond time-resolved regimes, resulting in an ultrafast optical technique to probe the dynamics of various excitations. Recent work on stimulated Raman and pump-probe spectroscopies showed the enhanced resolution via photon entanglement through the model calculations \cite{Dorfman_JCP2021,Dorfman_PRA2016,Swidzinsky_2021}. Nevertheless, quantum states of light may be powerful for exploiting the ultrafast electronic processes in semiconductor materials. This has been indicated from the semiconductor quantum-light sources taking the advantage of generating high photon flux and the scalability \cite{Schmidt_NatCommun2018,Shields_NatPht2007}. 
%The temperature-controlled SPDC for generating entangled photon pairs may yield the two-photon absorption signal taking the advantage of adjustable bandwidth and time delay between the photons \cite{URen_PRL2019}. 
%The temperature-controlled SPDC for generating entangled photon pairs has been demonstrated to yield the two-photon absorption signal overcoming the difficulties one encounters: the redundant work of using hundreds of nonlinear crystals with different lengths and the requirement of prior knowledge of the intermediate level in the absorbing matter. 
More delicate information about the structure and dynamics in metal nanostructures and low-dimensional materials can be read out from the photon correlation \cite{Lian_Science2015,Atwater_NatMater2020,Yao_NatPhys2016,Zhang_JCP2018,Wang_PRL2015,Zhang_JPCL2019}. Subsequent theoretical proposals aiming at ultrafast multi-photon coincidence counting spectrum revealed the highly-resolved excitonic relaxation pathways and charge fluctuation through manipulating the temporal and spectral gates, not accessible by classical pulses \cite{Zhang_JCP2018,Asban_PNAS2019,Ye_PNAS2019}. %It turns out the unusual bandwidth properties of entangled photons, which form the basis for the developments of novel ultrafast optical spectroscopy including two-dimensional electronic spectroscopy and harmonic generation signals \cite{Mukamel_PRA2010,Ming_PRL1997,Miller_NatChem2014,Stone_Science2009}.
It turns out the unusual band spectrum of entangled photons %which forms the basis of the 
responsible for the superresolution nature beyond Heisenberg conjugation renders the quantum-light spectroscopy as a new promising route towards novel applications in ultrafast and ultra-sensitive tomography \cite{Mukamel_RMP2016,Ming_PRL1997,Miller_NatChem2014,Stone_Science2009,Saleh_PRL1998,Liu_CPL2021}. 
%Regardless of some experimental concerns already raised in the nonlinear spectroscopy using quantum states of photons, quantum-light spectroscopy has been widely considered as a new promising route towards novel applications in ultrafast and ultra-sensitive tomography \cite{Mukamel_RMP2016,Saleh_PRL1998,Liu_PRL2019,Liu_CPL2021}. This is owing to the superresolution nature beyond Heisenberg conjugation. 

In this Letter, we demonstrate new capabilities brought out by the use of entangled photons in femtosecond time-resolved coherent Raman spectroscopy, to monitor the electronic and vibrational coherences. This leads to the QFRS (Quantum femtosecond Raman spectroscopy). Specifically, we show the time-frequency resolutions, obtained by the photon entanglement. Explicit results are presented for a quantum extension of FAST CARS (Femtosecond adaptive spectroscopic technique for coherent anti-Stokes Raman scattering) \cite{Scully_PNAS2002,Scully_PNAS2007,Pestov_Science2007}. Extensive works have been done in the population relaxation with entangled-photon spectroscopy \cite{Dorfman_PRA2016,Dorfman_JPCL2014}. The ultrafast dynamics of electronic coherence fundamentally important for reaction kinetics and internal conversions, however, is a challenging task due to its rapid decay \cite{Musser_NatPhys2015,Seel_JCP1991,Zhang_SR2016,Zhou_NatCommun2019,Fleming_Nature2007,Kowalewski_PRL2015}. We show how the entangled-photon spectroscopy could be useful here.

%In this Letter, we develop a femtosecond time-resolved coherent Raman spectroscopy using entangled photons that gives a highly-resolved real-time tomography of the electronic coherence and its dephasing. This leads to the QFRS (Quantum femtosecond Raman spectroscopy), as a novel extension of FAST CARS (Femtosecond adaptive spectroscopic technique for coherent anti-Stokes Raman scattering) commonly used to probe vibrational coherence and to optimize the efficiency of optical signals \cite{Scully_PNAS2002,Scully_PNAS2007,Pestov_Science2007}. While the population relaxation of molecular excited states may be visualized by  entangled-photon spectroscopy \cite{Dorfman_PRA2016,Dorfman_JPCL2014}, the ultrafast dynamics of electronic coherence fundamentally important for reaction kinetics and internal conversions is still a challenging task due to its rapid decay \cite{Musser_NatPhys2015,Seel_JCP1991,Zhang_SR2016,Zhou_NatCommun2019,Fleming_Nature2007,Kowalewski_PRL2015}. 

\begin{figure*}[t]
 \captionsetup{justification=raggedright,singlelinecheck=false}
\centering
\includegraphics[scale=0.76]{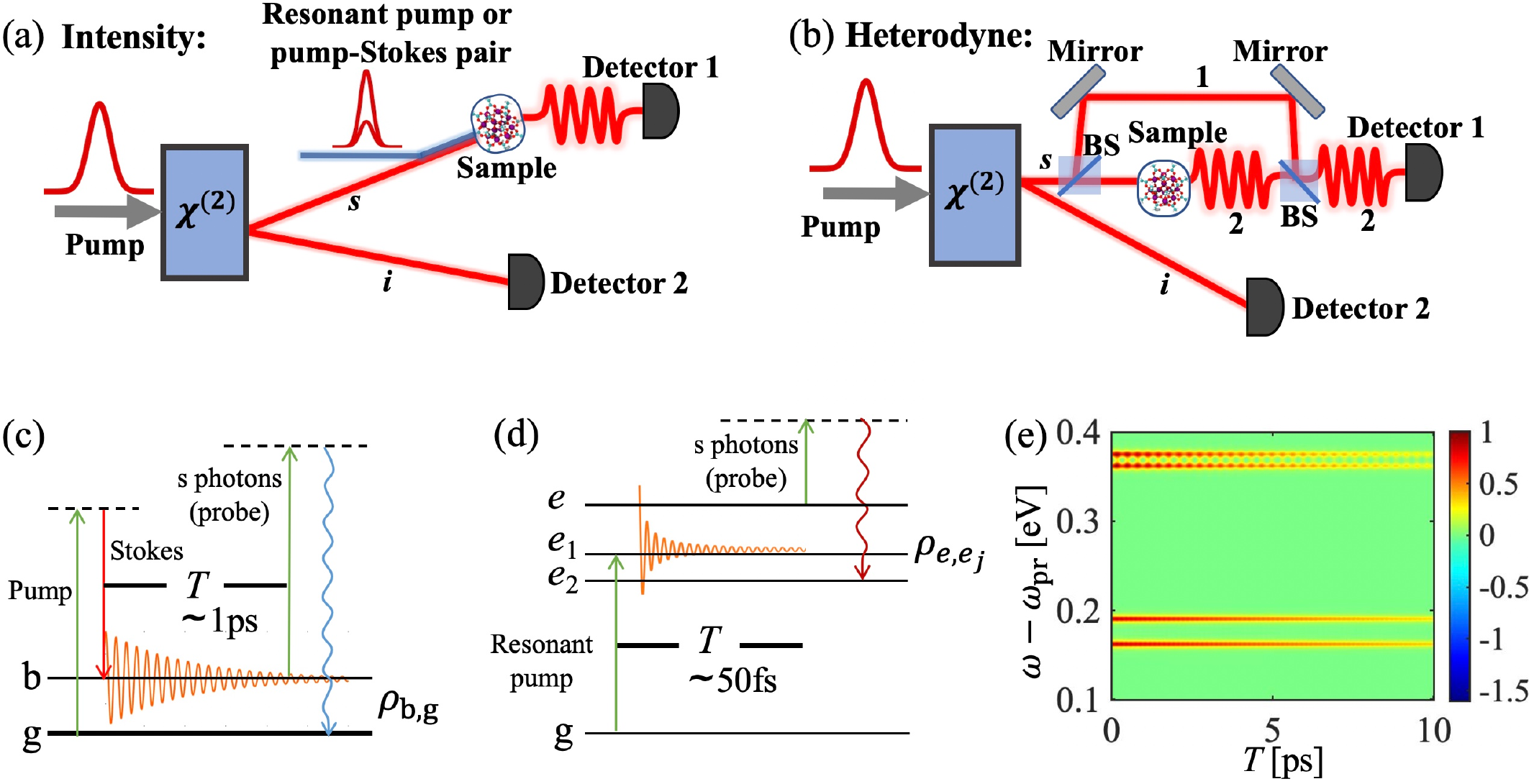}
\caption{(a) Schematic of entangled twin photons as ultrafast probe for molecules, where the parametric down conversion through a beta baruim borate (BBO) crystal and multi-photon detection are presented. (b) Quantum heterodyne scheme of detection where a local oscillator interferes with emitted photons in s arm; Detector 1 records the interference term whereas detector 2 records the idler photon number, yielding the joint detection. (c) Level scheme of microscopic model in Quantum FAST CARS. (d) Level scheme of microscopic model in QFRS for electronically excited states. (e) Quantum FAST CARS signal from Eq.(\ref{SQC}), taking the 4 Raman-active modes $\text{A}_1$, $\text{E}$ and $\text{T}_2$ in methane ($\mathrm{CH}_4$).}
\label{F1}
\end{figure*}

{\it Quantum FAST CARS}.--To start off, we consider the FAST CARS with entangled photons, where the molecules are driven by a pair of classical pulses. After a time delay, the beam in s arm of the entangled twin photons serves as a Raman probe off-resonantly interacting with molecules, whereas the idler beam propagates freely and provides a reference, as depicted in Fig.\ref{F1}(a). Referring to the level scheme in Fig.\ref{F1}(c), the field-molecule interaction of Raman process is
\begin{equation}
    \begin{split}
        V(t) = \sum_{j=1}^N\sum_{b} \alpha_{bg}^{(j)} |b\rangle\langle g|_j E_s(t) E_s^{\dagger}(t) e^{i(\omega_b-\omega_g)t} + \text{h.c.}
    \end{split}
\label{RamanIn}
\end{equation}
where $\alpha_{bg}^{(j)}$ is the Raman polarizability and the field $E_s(t)$ contains multiple frequency modes \cite{SM}. $N$ denotes the number of molecules. $|\Psi\rangle = \frac{1}{\sqrt{{\cal N}}}\int_{-\infty}^{\infty}\int_{-\infty}^{\infty}\text{d}\omega_s\text{d}\omega_i\ \Phi(\omega_s,\omega_i)|1_{\omega_s},1_{\omega_i}\rangle$ and
\begin{equation}
    \begin{split}
        \Phi(\omega_s,\omega_i) = E_0(\omega_s+\omega_i)\text{sinc}\left[\frac{\Delta k(\omega_s,\omega_i)L}{2}\right]e^{i\Delta k(\omega_s,\omega_i)L/2}
    \end{split}
\label{Phi}
\end{equation}
is the wave function for the entangled twin photons with $\Delta k(\omega_s,\omega_i)L=(\omega_s-\frac{\omega_0}{2})T_s+(\omega_i-\frac{\omega_0}{2})T_i$ where $T_s (T_i)$ is the time delay between the photons in $s$ (reference) arm and the SPDC pump field, due to the group velocity dispersion in the nonlinear crystal. ${\cal N}$ is a normalization factor.

A joint detection of the spectral-resolved transmissions in two arms gives the intensity-correlated signal $S(\omega,\omega_i;T)=\langle E_s^{\dagger}(\omega)E_i^{\dagger}(\omega_i)E_i(\omega_i)E_s(\omega)\rangle_{\rho}$ where $E_s(\omega)$ and $E_i(\omega_i)$ denote the Fourier component of the electric fields in s and reference arms, respectively. $T$ is the time delay of $s$-arm photons relative to the pump pulse. Using the perturbation expansion against Eq.(\ref{RamanIn}), we find the 8-point field correlation function playing a key role in understanding the unusual spectroscopic properties of Raman signals with photon entanglement. %where $E_s(\tau)=\frac{1}{2\pi}\int E_s(\omega)e^{-i\omega (\tau-T)}d\omega$ and $E_s(\omega)={\cal E}_{s,\omega} a_{s,\omega}$. ${\cal E}_{s,\omega}=\sqrt{2\pi\omega_s/\text{V}}$ and $a_{s,\omega}$ is the annihilation operator of photon mode in $s(i)$ arm. $T$ is the time delay acquired by the $s$-arm photons to the pump pulse. 
A lengthy algebra leads to the Quantum FAST CARS signal
\begin{equation}
    \begin{split}
        S_{\text{QV}}(\omega,\omega_i;T) = & \frac{N(N-1)}{8\pi^2{\cal N}} |{\cal E}_{s,\omega}|^6 |{\cal E}_{i,\omega_i}|^2 \bigg| \sum_b \alpha_{bg}^* \rho_{bg}(T) \\[0.15cm]
        & \qquad\qquad \times \Phi(\omega-\omega_{bg}-i\gamma_{bg},\omega_i ) \bigg |^2
    \end{split}
\label{SQC}
\end{equation}
with $\rho_{bg}(t)=\rho_{bg}e^{-(i\omega_{bg}+\gamma_{bg})t}$ where $\gamma_{bg}^{-1}$ quantifies the dephasing of the vibrational coherence. ${\cal E}_{s(i),\omega}=\sqrt{2\pi\omega_{s(i)}/\text{V}}$ and $\text{V}$ is bulk volume. Besides the enhanced signal intensity via coherence, Eq.(\ref{SQC}) indicates the spectral resolution governed by photon entanglement that contains new control knobs such as photon arrival time and bandwidth.

\begin{figure*}[t]
 \captionsetup{justification=raggedright,singlelinecheck=false}
\centering
\includegraphics[scale=0.76]{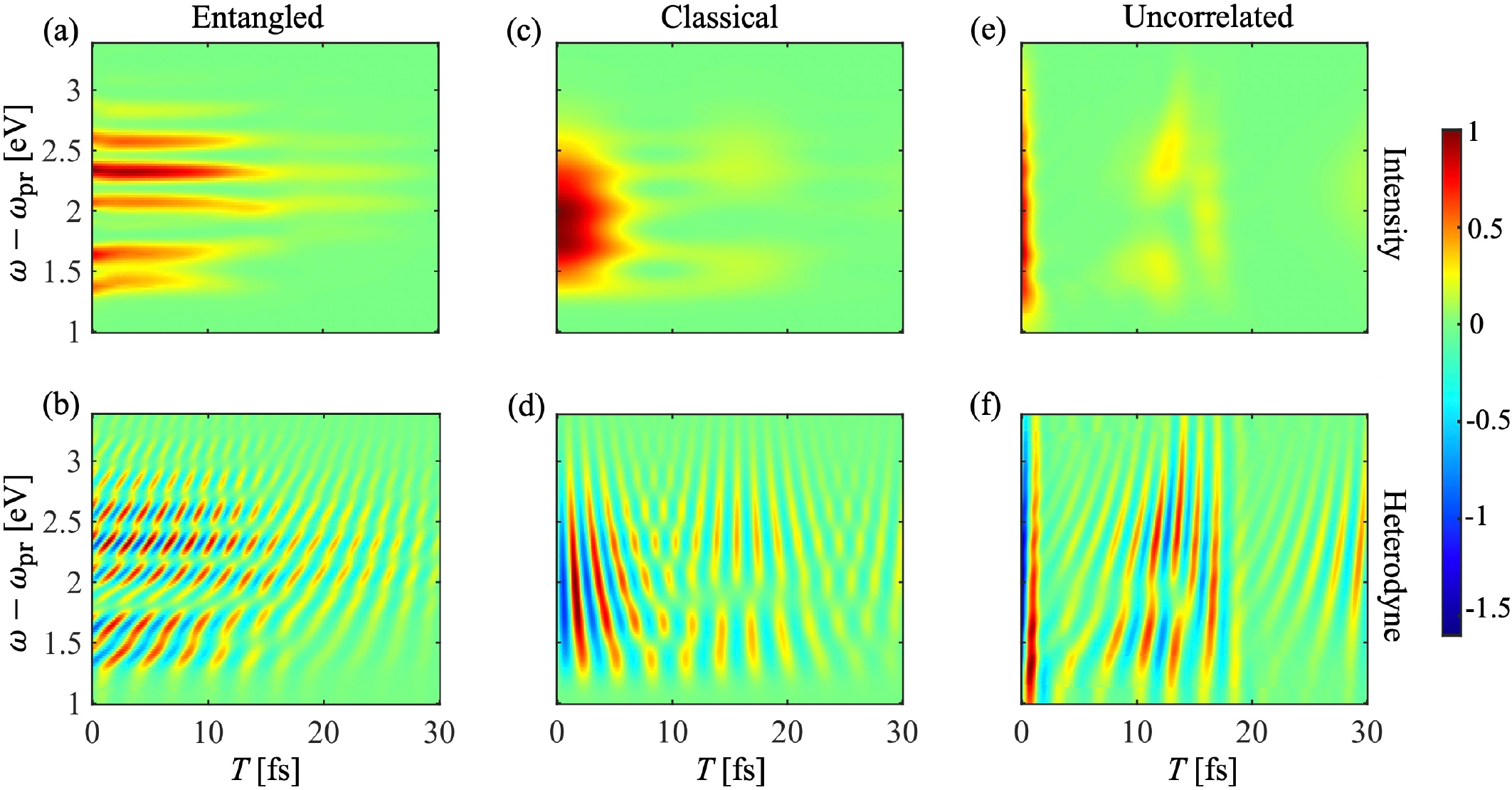}
\caption{(First column) (a) Intensity-correlated QFRS for time-evolving electronic coherence versus the delay $T$ between entangled photons and resonant pump pulse; (b) Same as (a) but for heterodyne detection. (c,d) Same as (a,b) but using classical pulse as probe having bandwidth $\sigma_0=0.82$eV. (e,f) Same as (a,b) but using uncorrelated separable state of photons with the bandwidth $\sigma_0=0.82$eV ($\sigma_0^{-1}=5$fs). Parameters are taken from 4-oriented amino-4’nitrostilbene \cite{Rouxel_PRL2018}, i.e., $\omega_e=7.1$eV, $\omega_{e_1}=5.3$eV, $\omega_{e_2}=5.7$eV, $v_h=0.26$eV, $F_1=2.2,\ F_2=1.3$, $D_1^{-1/2}=30$fs, $D_2^{-1/2}=20$fs and $T_s=T_i=30$fs.}  
\label{RamanS2D}
\end{figure*}

For a neat understanding of the signal, we focus on the optimal photon entanglement, i.e., $T_s=T_i$. The two-photon amplitude in Eq.(\ref{Phi}) indicates a Raman resonance $\omega-\omega_{\text{pr}}=\omega_{bg}$ defining the probe frequency $\omega_{\text{pr}}=\omega_0-\omega_i$. This leads to the spectral resolution $T_s^{-1}$, which reveals a considerable enhancement not accessible by classically shaped pulses. To make this clear, we calculate the Raman signals with two incoherent states of light: (i) Fock state of photons that is fully separable and (ii) the psuedo-thermal light $\rho_{\text{pht}} = \frac{1}{{\cal N}}\iint\text{d}v \text{d}v_i |\Phi(v,v_i)|^2 |1_{s,v};1_{i,v_i}\rangle\langle 1_{s,v};1_{i,v_i}|$ having classical correlation which has been used in information processing. For (i), the spectral resolution is governed by the single-photon amplitude $\Phi_s(\omega_s-\omega_0)\propto e^{-(\omega_s-\omega_0)^2/2\sigma_0^2}$ having the identical time duration $\sigma_0^{-1}$ as the pump pulse, as seen from Eq.(16) in Ref.\cite{SM}. In CARS experiments, the conjugated time and frequency resolutions results in the challenge to distinguish the vibrational modes once $|\omega_{b_1,b_2}|\ll \sigma_0$. Using photon entanglement, the spectral resolution can be controlled independently and can be therefore greatly enhanced such that $T_s^{-1} < |\omega_{b_1,b_2}| \ll \sigma_0$, evident by Fig.\ref{F1}(e) that keeps high temporal resolution. For (ii) as entailed by Eq.(15) in Ref.\cite{SM}, the signal shows a high spectral resolution $T_s^{-1}$, whereas the temporal resolution is no longer there. This is expected from the random phase between photon pairs that results in arbitrary arrival time of photons.

{\it Intensity-correlated QFRS}.--To monitor the ultrafast dynamics of electronic coherence in excited states of molecules, the probe pulse has to act after a delay $T$ following a resonant photoexcitation or a propagation of the excited states, depicted in Fig.\ref{F1}(d). The Raman signal is produced when the s-arm photon scatters off the excited-state coherence resulting in the electronic polarizability $\alpha(t)$ describing the Raman transitions between excited states. Proceeding via the algebra similar as before, we find the QFRS signal \cite{SM}
\begin{equation}
    \begin{split}
        S_{\text{QE}}(\omega,\omega_i;T) = \frac{N(N-1)}{32\pi^4{\cal N}}|{\cal E}_{s,\omega}|^6 |{\cal E}_{i,\omega_i}|^2 \bigg|\sum_{e'} \alpha_{e,e'}^*f_{e,e'}(T)\bigg|^2
    \end{split}
\label{SF}
\end{equation}
where the Raman line-shape function is defined as
\begin{equation}
    \begin{split}
        f_{e,e'}(T) & = \int_{-\infty}^{\infty}\text{d}\tau \int_{-\infty}^{\infty}\text{d}\omega' \rho_{e,e'}(\tau) e^{i(\omega-\omega')(\tau-T)}\Phi(\omega',\omega_i).
    \end{split}
\label{lineshape}
\end{equation}
Eq.~(\ref{lineshape}) indicates the role of the photon entanglement whose unusual band properties may provide versatile tool for controlling the ultrafast electron dynamics in molecules.

To elaborate this, the electronically excited-state dynamics coupled to nuclear motions has to be considered. This may facilitate a quantitative understanding of the inhomogeneous dephasing in realistic materials. %e.g., molecules embedded in dense medium that is responsible for rich effects. 
We adopt the Fr\"ohlich-Holstein model for molecules
\begin{equation}
    \begin{split}
        H = H_{\text{vib}}|g\rangle\langle g| + \sum_i \left(\omega_{e_i,g} + \lambda_s^{(i),2}v_s + H_{\text{vib}}^i\right)|e_i\rangle\langle e_i|
        %H = \sum_j \tilde{\omega}_{e_j,g}|e_j\rangle\langle e_j| + \sum_s v_s b_s^{\dagger}b_s - \sum_{j,s} \lambda_s^{(j)}v_s |e_j\rangle\langle e_j| \left(b_s + b_s^{\dagger}\right)
    \end{split}
\label{Hm}
\end{equation}
and $H_{\text{vib}}=\sum_s v_s b_s^{\dagger}b_s$ and $H_{\text{vib}}^i=\sum_s \left[v_s b_s^{\dagger}b_s - \lambda_s^{(i)} v_s (b_s + b_s^{\dagger})\right]$, where $b_s$ is the annihilation operator of the $s$-th vibration having the frequency $v_s$. $\lambda_s^{(i)}$ quantifies the vibronic coupling for the $i$-th electronic state. Propagating the nuclear wave packets and using Eq.(\ref{SF}) and (\ref{lineshape}), the intensity-correlated QFRS signal can be found
\begin{equation}
    \begin{split}
        & S_{\text{QE}}(\omega,\omega_i;T) = \frac{N(N-1)}{8\pi^2{\cal N}}|{\cal E}_{s,\omega}|^6 |{\cal E}_{i,\omega_i}|^2\\[0.15cm]
        & \qquad\qquad \times \bigg|\sum_{e_1}\sum_{n=0}^{\infty} \alpha_{e,e_1}^* g_{n,e_1}(\omega,T)\rho_{e,e_1}^{(n)}(T)\bigg|^2
    \end{split}
\label{SB}
\end{equation}
where $g_{n,e_j}(\omega,T) = \int_{-\infty}^{\infty}\text{d}\tau\ e^{i(\omega-\tilde{\omega}_{e,e_j}-n v_h)\tau} e^{-D_j(\tau^2+2T\tau)}\tilde{\Phi}(\tau,\omega_i)$ and $\tilde{\Phi}(\tau,\omega_i)\equiv\frac{1}{2\pi}\int_{-\infty}^{\infty}\Phi(\omega,\omega_i)e^{-i\omega \tau}\text{d}\omega$ \cite{SM}. $\rho_{e,e_j}^{(n)}(t)/\rho_{e,e_j}(0) = S_{n,j}e^{-i(\tilde{\omega}_{e,e_j}+nv_h)t-D_j t^2}$ is the vibronic coherence with $n$ harmonics of the high-frequency vibrations and $S_{n,j}=e^{-F_j}F_j^n/n!$ is the Franck-Condon factor. $D_j=m(\lambda_l^{(j)}-\lambda_l^{(e)})^2$ and $F_j=(\lambda_h^{(j)}-\lambda_h^{(e)})^2$ quantify the coupling of excitons to low- and high-frequency vibrations, respectively \cite{Barbara_JPC1996}. The superposition of vibronic coherences yields the electronic coherence, i.e., $\rho_{e,e_j}(t)=\sum_{n=0}^{\infty}\rho_{e,e_j}^{(n)}(t)$.

Notably, for the timescale $T<1/\sqrt{D_j}$, we can find the line-shape function for arbitrary two-photon amplitude, namely, $g_{n,e_j}(\omega,T) \simeq \Phi(\omega-\tilde{\omega}_{e,e_j}-n v_h+2iD_j T,\omega_i)$. Eq.(\ref{SB}) indicates the multiple Raman resonance $\omega-\omega_{\text{pr}}=\tilde{\omega}_{e,e_1}+n v_h$ in which the intensity is governed by the Franck-Condon factor quantifying the vibronic interaction. The spectral resolution is $T_s^{-1}$ that may be greatly enhanced, as dictated by the two-photon amplitude. The Raman signals in Eq.(\ref{SB}) are highly time- and frequency-resolved, revealing the ultrafast coherent dynamics of the electronically excited states.

\begin{figure}[t]
 \captionsetup{justification=raggedright,singlelinecheck=false}
\centering
\includegraphics[scale=0.42]{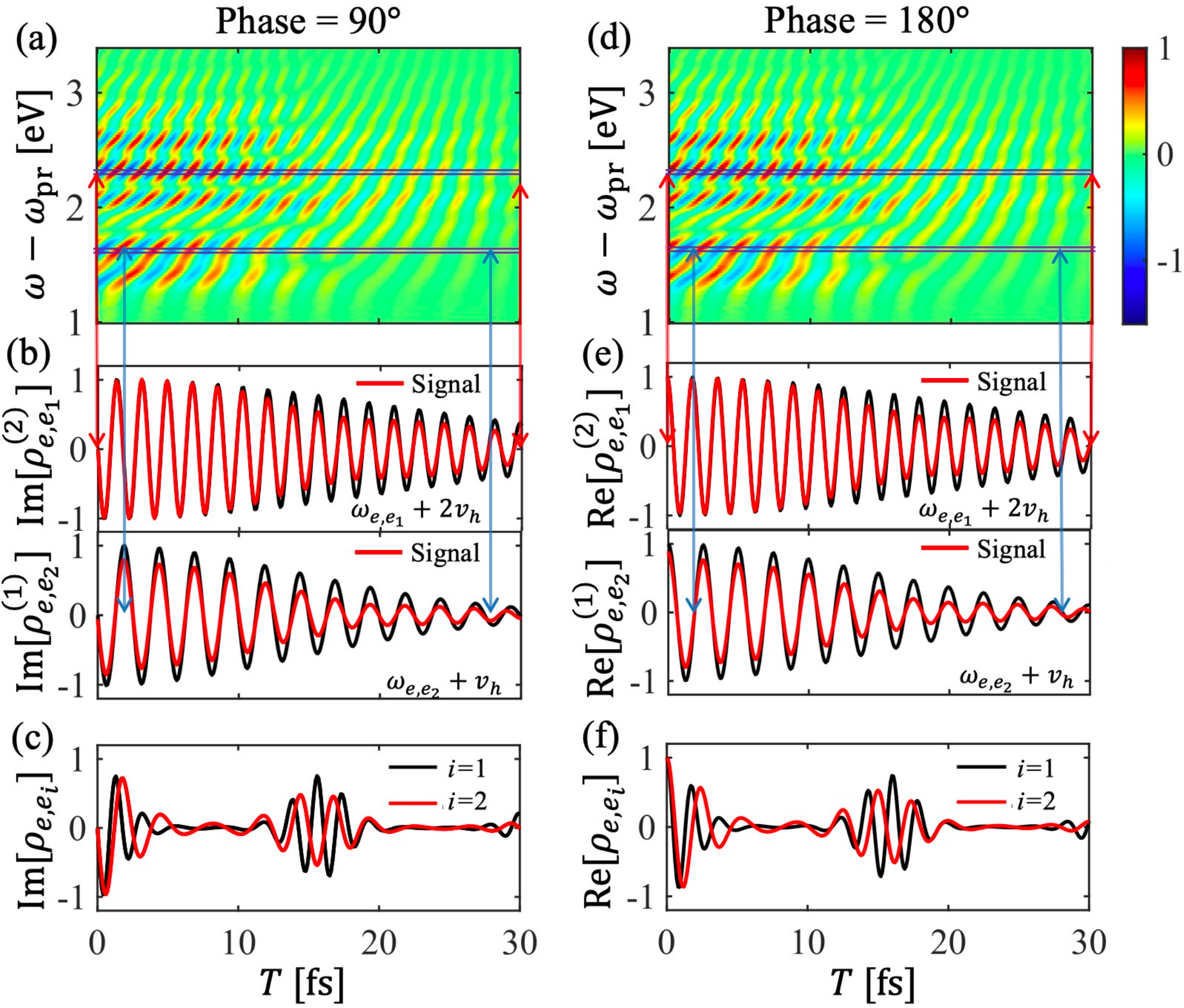}
\caption{Heterodyne-detected QFRS and dynamics of vibronic coherence. (Left column) (a) Heterodyne-detected QFRS with the phase associated with $\Phi(\bar{\omega},\omega_i)$ being $-\pi/2$ (a certain phase of local oscillator). (b, up) Red line is for the 1D slice of (a) at $\omega-\omega_{\text{pr}}=\omega_{e,e_1}+2v_h=2.32$eV; Black line is for the imaginary part of $n=2$ harmonics of the vibronic coherence $\rho_{e,e_1}$, i.e., $\text{Im}[\rho_{e,e_1}^{(2)}]$. (b, down) Same as (b, up) but for $\omega-\omega_{\text{pr}}=\omega_{e,e_2}+v_h=1.66$eV and imaginary part of $n=1$ harmonics of the vibronic coherence $\rho_{e,e_2}$. (c) Imaginary part of time-evolving vibronic coherence, i.e., $\text{Im}[\rho_{e,e_1}]$ (black) and $\text{Im}[\rho_{e,e_2}]$ (red). (Right column) (d) Same as (a) but for the phase associated with $\Phi(\bar{\omega},\omega_i)$ being $-\pi$. (e) Same as (b) but for 1D slices of (d) (red lines) and $\text{Re}[\rho_{e,e_1}^{(2)}],\ \text{Re}[\rho_{e,e_2}^{(1)}]$ (black lines). (f) Same as (c) but for $\text{Re}[\rho_{e,e_1}]$ and $\text{Re}[\rho_{e,e_2}]$. Parameters are the same as Fig.~\ref{RamanS2D}.}  
\label{SHDcoh}
\end{figure}

%For a more explicit manner, we simulate the QFTRS in Fig.~\ref{RamanS2D}. by taking the parameters from 4-oriented amino-4’nitrostilbene as given in Fig.~\ref{RamanS2D}. %i.e, $\omega_e=7.1$eV, $\omega_{e_1}=5.3$eV, $\omega_{e_2}=5.7$eV, $v_h=2100$cm$^{-1}$, $F_1=2.2,\ F_2=1.3$, $D_1^{-1/2}=30$fs and $D_2^{-1/2}=20$fs. The pulse parameters are adjustable, so that we can properly assume $\sigma_0^{-1}=5$fs, $T_s=30$fs. 
Fig.~\ref{RamanS2D}(a) illustrates the intensity-correlated QFRS, where the electronic coherence $\rho_{e,e_i};i=1,2$ is monitored, yielding the Raman resonance corresponding to the two electronically excited states at $\omega_{e_1}=5.3$eV and $\omega_{e_2}=5.7$eV. Using the short pulse, we can readily resolves the vibronic couplings. In particular, the signal contains the Raman resonance $\omega-\omega_{\text{pr}}=\omega_{e,e_1}+nv_h\ (n=1,2,3,4)$ and  $\omega_{e,e_2}+mv_h\ (m=0,1,2)$, where the peaks at $\omega_{e,e_1}+nv_h$ decay slower and dominate at longer time delay $T$. This is because of the weaker influence from the low-frequency vibrations leading to slower dephasing of electronic coherence. It is worth noting that only a few harmonics of the vibrational modes can be seen from the Raman signal for electronic coherence. This results from the Franck-Condon effect yielding a Poissonian distribution of the vibronic energy seen from $S_{\text{QE}}(\omega,\omega_i;T=0)$. 
%evident by signal $S_{\text{HM}}$ in Eq.~(\ref{SB}) at $T=0$.
Fig.~\ref{RamanS2D}(c) illustrates the intensity-correlated QFRS with a classical probe pulse. The broadband pulse has a high temporal resolution, but the individual electronically excited states, as well as their coupling to vibrations are poorly resolved. This has been a long existing bottleneck in ultrafast Raman spectroscopy. 

We further examine the Fock state of photons that is fully separable with neither correlation nor entanglement, in Fig.~\ref{RamanS2D}(e). It shows that the fully separable Fock photon state yields as poor resolutions as the Raman signals with classical probe. This is because the field correlation function is factorized into a product of field intensities.

It is worth noting that the intensity-correlated QFRS may smear out the phase information of the excited states, as seen from the spectral-resolved condition $T_s^{-1}<|\omega_{e_1,e_2}|$ so that $|\rho_{e,e'}|^2$ dominates the signal. A new spectroscopic signal is therefore essential to be exploited, as will be entailed next.

{\it Heterodyne-detected QFRS}.--We let the s-arm photons serve as a local oscillator interfering with the emitted photons, as depicted in Fig.\ref{F1}(b). The field directed into detector 1 carries the superposition state $|e\rangle+e^{i\phi}|e'\rangle$ assuming 50/50 beam splitters. The joint detection of photons with anti-Stokes shift yields the measurement $\propto \text{Re}(e^{-i\phi}\rho_{e,e'})$. In such a spirit, we find the heterodyne-detected signal explicitly \cite{SM} %$S_{\text{HD}}(\omega,\omega_i;T) = 2\text{Im}\left[\int_{-\infty}^{\infty}\text{d}t e^{i\omega(t-T)}\langle E_i^{\dagger}(\omega_i)E_s^{\dagger}(\bar{\omega})E_s(t-T)\alpha(t) E_i(\omega_i)\rangle\right]$.
\begin{equation}
    \begin{split}
        S_{\text{QEHD}}(\omega,\omega_i;T) = \frac{N|{\cal E}_{s,\omega}|^2 |{\cal E}_{i,\omega_i}|^2}{\pi{\cal N}} \text{Im}\bigg[\sum_{e'}\alpha_{e,e'}^*\Phi^*(\bar{\omega},\omega_i)f_{e,e'}(T)\bigg]
    \end{split}
\label{SHDe}
\end{equation}
with the same definition of Raman line-shape function and electronic polarizability as in Eq.(\ref{SF}) and (\ref{lineshape}).

For the molecules in condensed phases, we incorporate into the signal the nuclear dynamics interacting with electronically excited states. The heterodyne-detected QFRS reads
\begin{equation}
    \begin{split}
        S_{\text{QEHD}}(\omega,\omega_i;T) = & \frac{2N}{{\cal N}}|{\cal E}_{s,\omega}|^2 |{\cal E}_{i,\omega_i}|^2 \sum_{e_1}\sum_{n=0}^{\infty} \text{Im}\Big[\alpha_{e,e_1}^* \rho_{e,e_1}^{(n)}(T)\\[0.1cm]
        & \qquad\qquad \times \Phi^*(\bar{\omega},\omega_i) g_{n,e_1}(\omega,T)\Big]
    \end{split}
\label{SBh}
\end{equation}
where $g_{n,e_j}(\omega,T)$ is of the identical definition as in Eq.(\ref{SB}). From Eq.(\ref{SBh}), $S_{\text{HD}}(\omega,\omega_i;T)$ visualizes the full transient coherence dynamics including phase and intensity, via the amplitude $\Phi(\bar{\omega},\omega_i)$ whose phase varies with the optical path.

To track the fast coherence dynamics in further, we plot the heterodyne-detected QFRS in Fig.~\ref{RamanS2D}(b), which unambiguously reveals the oscillations. The heterodyne signal predicts the same Raman resonance as the signal in Fig.~\ref{RamanS2D}(a). The advantage of photon entanglement is shown in Fig.~\ref{RamanS2D}(d) which produces a much worse spectral resolution with a classical broadband probe, despite of the temporal resolution. This is extensively supported from the comparison to the heterodyne signal using Fock state of photons, as depicted in Fig.~\ref{RamanS2D}(e) where neither correlation nor entanglement has been involved. 
%which produces the heterodyne-detected signal with classical probe pulse. The results reveal a much worse spectral resolution with a broadband pulse, despite of the high temporal resolution.

Fig.~\ref{SHDcoh} shows the heterodyne-detected QFRS with different phases of the local oscillator field that leads to a global phase in $\Phi(\bar{\omega},\omega_i)$. It can be seen from Fig.~\ref{SHDcoh}(a) and~\ref{SHDcoh}(d) the fast oscillations that selectively track the real-time dynamics of the coherence phase (imaginary and real parts), given the high spectral resolution revealing the vibronic states. As time evolves, Fig.~\ref{SHDcoh}(b) and~\ref{SHDcoh}(e) illustrate the Fourier components of the vibronic coherence (imaginary and real parts) resolved from the oscillations with various frequencies, e.g., $\omega_{e,e_1}+2v_h$ and $\omega_{e,e_2}+v_h$ as a result of the vibronic coupling. This leads to the global beating feature of the vibronic coherence, depicted in Fig.~\ref{SHDcoh}(c) and~\ref{SHDcoh}(f) revealing the phase information. %Therefore, the quantum heterodyne detection provides additional knobs for exploiting the ultrafast coherent dynamics of electrons.

%We now elaborate the role of photon entanglement essential for improving the resolutions of the time-resolved Raman spectroscopy. We examine the Fock state of photons which is fully separable such that $\Phi(\omega_s,\omega_i)=\Phi_s(\omega_s-\omega_0)\Phi_i(\omega_i-\omega_0)$ with $\Phi_v(\omega_v-\omega_0)=A_v/(\omega_v-\omega_0+i\sigma_0);\ v=s,i$ and $\omega_0$ matches the central frequency of the probe pulse used in the classical Raman signal shown in Fig.\ref{RamanS2D}(c) and \ref{RamanS2D}(d). Such state has no correlation and zero photon number fluctuation that has been used in damage-free detection and quantum simulation \cite{Ourjoumtsev_PRL2006,Wolf_NatCommun2019,Yamamoto_PRL2001,Wang_NSR2021}. Fig.~\ref{RamanS2D}(e) and~\ref{RamanS2D}(f) show the fully separable Fock state of photons yields the same resolution as the classical Raman spectroscopy in Fig.~\ref{RamanS2D}(c) and~\ref{RamanS2D}(d) that are poorly resolved with short pulses, for both homodyne and heterodyne schemes of detection. This is because the field correlation function is factorized into a product of field amplitudes. Derivations of the ultrafast Raman signals incorporating the Brownian oscillator model %for entangled twin state, classical and Fock states are given in SM.

{\it Experimental feasibility}.--Since the experiments using electronically ground states are easier and have been accomplished, we will focus on the feasibility of Quantum FAST CARS. The vibrational coherence is scattered off by the s photons entangled with the idler photons having a delay from the Stokes pulse. Dramatically different from the FAST CARS, the probe pulse using entangled twin photons can be broadband, rather than the narrow-band one. %e.g., femtosecond pulse giving a high temporal resolution, due to the time and frequency not conjugated with each other. 
Typically, the time duration of fast probe is taken to be $\sigma_0^{-1}=35$fs and the delay controlled by the crystal length is $T_s=1$ps yielding the Raman signal with frequency resolution of 33cm$^{-1}$ sufficient for a variety of organic molecules. This allows to monitor the fast-evolving vibrational coherence, provided the same spectral resolution as the FAST CARS.

%This allows a time delay $\sim 300$fs between the probe and Stokes pulses, considerably shorter than the $>1$ps acquired by FAST CARS using picosecond pulse as probe \cite{Pestov_Science2007}. Therefore a double enhancement of Raman signal intensity may be expected, as $e^{-300/1000}/e^{-1}\simeq 2$ when using entangled twin photons, provided the same spectral resolution as the FAST CARS.

We employ the methane ($\mathrm{CH}_4$) which is a ubiquitous molecule with well characterized spectral lines. The Raman-active transition $\mathrm{A}_1$ at 2914$\mathrm{cm}^{-1}$ of methane, for instance, can be probed with focusing the femtosecond pump and Stokes pulses followed by s photons.
%The number density inside a methane cell is $\sim 2.5\times 10^{15}/\mathrm{mm}^3$. 
The s photon pulse intensity is of $10^{-9}$ magnitude lower than the classical pulse. %As compared to the classical pump, the intensity of s photon probe is $\sim 10^{-9}$ lower. 
In a typical CARS scheme the electronic transition dipole is $\sim 1$ debye with pump detunings on the order of $10^5$ $\mathrm{cm}^{-1}$, the quantum light signal is only $\sim 10^{-1}/\mathrm{s}$. Nevertheless, approaching resonance by reducing detuning with 2 orders, may give an extra 4 orders enhancement of the signal without laser damage. One can then expect the quantum FAST CARS signal at a level of ~$10^3/\mathrm{s}$, which is sufficient for the detection. We note here that the quantum FAST CARS scheme possesses an inherent advantage due to the multi-photon coincidence counting. This plays an essential role in increasing the signal-to-noise ratio for quantum FAST CARS. 

{\it Summary}.--We developed a new time- and frequency-resolved Raman spectroscopy using entangled photons and interferometry, as a quantum extension of FAST CARS. Substantially, QFRS was further proposed to probe the electronic coherence. %This technique measures the frequency-resolved Raman scattering of a quantum-field probe as a function of time delay $T$ relative to the pump pulse. 
In contrast to existing paradigms \cite{Mukamel_RMP2016,URen_PRL2019,Dorfman_JPCL2014}, %such as the stimulated Raman spectroscopy, 
QFRS is sensitive to the electronic coherence only and is population-free, making it uniquely suitable for monitoring the electronically excited state dynamics during a short timescale. %which may survive during a short timescale. 
We developed a microscopic theory for the signals involving the intensity- and heterodyne-detected modes. These demonstrated that QFRS can measure the time-varying energy gap between the vibronic states with a high spectral resolution. The rapid oscillation and decay of electronic coherence can be fully visible in the time-resolved spectrum. %where the latter contains the information about the low-frequency molecular vibrations. 
Such temporal- and spectral-resolved nature is not attainable in conventional Raman spectra. The QFRS, as a new coherent Raman technique, offers new perspective for the ultrafast dynamics in photophysical systems including low-dimensional semiconductor materials, exciton polaritons and nano-plasmonics.

%Finally, we discussed the experimental feasibility of quantum FAST CARS. It would be prospective in near future from a back-of-the-envelop analysis taking into account the experimental parameters.

\vspace{0.15cm}
Z.D.Z. and T.P. contributed equally to this work.

\vspace{0.15cm}
We gratefully thank Renbao Liu from CUHK for the instructive discussions and suggestions. Z.D.Z. gratefully acknowledges the support of ARPC-CityU new research initiative/infrastructure support from central (No. 9610505). T.P., G.S.A. and M.O.S. gratefully acknowledge the support of Air Force
Office of Research (Grant No. FA9550-20-1-0366), Office
of Naval Research (Grant No. N00014-20-1-2184), Robert A. Welch Foundation (Grant No. A-1261; A-1943), and National Science Foundation (Grant No. PHY-2013771).

\end{document}